\documentclass{kluwer}    

\newdisplay{guess}{Conjecture}

\def\BP{Ballesteros-Paredes}
\def\VS{V\'azquez-Semadeni}

\begin{document}

\begin{opening}         

\title{Fragmentation and Structure Formation}

\author{Enrique \surname{V\'azquez-Semadeni}}  
\runningauthor{Enrique V\'azquez-Semadeni}
\runningtitle{Fragmentation}
\institute{Centro de Radioastronom\'\i a y Astrof\'\i sica, UNAM}
\date{13 October 2003}

\begin{abstract}
I present an overview of the hierarchy of structures existing in the
interstellar medium (ISM) and the possible mechanisms that cause the
fragmentation of one level into the next, with the formation of
stars as its last step. Within this framework, I then give
an overview of the contributions to this session. Numerical work
addresses, at the largest scales, the shaping 
and formation of structures in the ISM through 
turbulence driven by stellar energy injection, and the resulting star
formation rate as a function of mean density. At the scales of molecular
clouds, results comparing observational and numerical data on the
density and velocity structure of turbulence-produced cores, as well
as their mass spectra, are summarized, together with existing
theories of core and star formation controlled by the
turbulence. Observationally, an attempt to discriminate between the
standard and turbulent models of star formation is presented, finding
inconclusive results, but suggesting that both turbulence and the
magnetic field are dynamically important in molecular clouds and their 
cores. Finally, various determinations of the magnetic field strength
and geometry are also presented.
\end{abstract}

\end{opening}

``Fragmentation'' is a very general term used in the study of star
formation (SF) and the interstellar medium (ISM) that describes the
sequential breakup of a diffuse and extended mass of gas (the ISM)
into ever smaller regions (clouds, clumps, cores), ultimately leading
to the formation of stars. Although a single term is used to
describe this iterative process, it most likely comprises a variety
of physical mechanisms operating at the various levels of the
sequence. The study of fragmentation thus represents an attempt to
acquire a unified view of the formation of structures in the ISM, with
star formation as its last step. A session on this topic therefore
naturally belongs in this conference. 



In order to properly discuss fragmentation, it is probably best to
first make an attempt at formulating practical definitions of what we
mean by this and a few other related terms. First, let us define a
``structure" within the ISM as a connected region having specific
values (or ranges of values) of the
physical variables, with at least some of them clearly distinguishable
from their values in the surroundings of the region. 
``Fragmentation" can then be defined as the formation of
small-scale structures within a larger-scale one. This 
immediately implies that fragmentation is a {\it multi-scale}
process, because it involves at least the scales of the ``parent''
and ``daughter'' structures. In particular,
interstellar clouds (either HI or molecular) are 
essentially high-density structures embedded within 
larger-scale, lower-density ones. It is of course possible that, 
by either the same mechanism that formed them, or a
different one, the small-scale structures fragment into even 
smaller structures, the process then being termed {\it 
hierarchical}, with many possible levels of fragmentation.

In the ISM, we can recognize a large hierarchy of structures
including, from larger to smaller scales, spiral arms (of sizes $L
\sim$ 10 kpc), 
``superclouds'' ($L \sim$ 1 kpc, $n \sim 1$ cm$^{-3}$) (Elmegreen \&
Elmegreen 1983), giant molecular clouds (GMCs) ($L \sim$ 50
pc, $n \sim 50$ cm$^{-3}$), ``individual'' molecular clouds (MCs)
($L \sim 5$ pc, $n 
\sim 10^3$ cm$^{-3}$), clumps ($L \sim$ 0.5 pc, $n \sim 10^4$ cm$^{-3}$)
and cores ($L \lesssim 0.1$ pc, $n > 10^4$ cm$^{-3}$) (see, e.g., the
reviews by Blitz 1993; Elmegreen 1993). Note, however,  
that, although it is only human to classify and categorize, the above
``categories'' really appear to merge into one another, the ISM most
likely really being a continuum (see the review by V\'azquez-Semadeni
et al. 2003 and references therein; also, the contribution by {\bf de
Avillez} in this session\footnote{In
this paper, I denote in {\bf boldface} the contributions to the present
session.}), with structures spanning all of the
aforementioned range of scales. This contrasts with the traditional
multi-phase view of the ISM (Field, Goldsmith \& Habing 1969; McKee \&
Ostriker 1977).


It is interesting to compare the
contents of this session with those of earlier reviews on the subject
(e.g., Scalo 1985; Elmegreen 1993). Such a comparison shows
that great progress has been made through the use of numerical
simulations of the turbulent ISM, allowing a deeper understanding and
an ever more quantitative 
determination of the role of stellar-driven turbulence in shaping the
ISM (see, e.g., the review by \VS\ (2002), and the contributions by
{\bf de Avillez} and {\bf Sarson et al.}), as well as the interiors of
molecular clouds, and in controlling various aspects of the SF 
process (see the review by {\bf \BP} and the
contribution by {\bf Klessen}). 


The physical mechanisms operating at the various hierarchical levels
of density structure formation in the ISM include a)
the gravitational (Jeans 1902), thermal (Field 1965; 
Field, Goldsmith \& Habing 1969; Pikel'ner 1968) and Parker (1966) linear 
instabilities, or combinations thereof (Elmegreen 1991,
1994); b) random compressions in the globally turbulent ISM, and
c) the nonlinear development of the instabilities themselves.

Specifically, the gravitational instability, parhaps aided by
cooling, is thought to be responsible for the formation of the
largest structures, or ``superclouds" (see, e.g., the review by
Elmegreen 1993), 
although the Parker instability may also contribute at these
scales (Franco et al. 2002). Within these, GMCs may be originated by
a) gravitational instability, again aided by cooling, within their parent
superclouds (e.g., Elmegreen 1993, 1994; Wada \& Norman 1999); b)
magneto-Jeans instabilities or modified swing amplification (e.g., Kim \& 
Ostriker 2001), and c) large-scale turbulent motions driven by
stellar activity (\BP, \VS\ \& Scalo 1999; de Avillez 2000;
{\bf de Avillez}; see also the review by Mac Low 
2002). Early cloudlet coagulation models (Oort 1954; Kwan 1979)
are apparently not anymore considered as a viable GMC formation
mechanism (Elmegreen 1990), as neither is the Parker instability ({\bf
Kim, Ryu \& Hong}). 

Regardless of their formation mechanism, GMCs, and the 
``individual" molecular clouds inside them, are supersonically
turbulent themselves (see, e.g., the review by Blitz 1993 and
references therein), while being magnetized and roughly isothermal,
with approximate equipartition between the turbulent and magnetic energies
(e.g., Myers \& Goodman 1988). Thus, numerical simulations of MHD isothermal 
turbulent flows should constitute good models of molecular clouds.
However, in this case, besides the structure-formation mechanism at work, 
an important parallel question is what is the driving mechanism
for the turbulence at every scale, and how ubiquitous it is.
This question arises because recent numerical results have shown that 
the turbulence, left undriven, decays rapidly, roughly in a
large-scale crossing time at the large-scale velocity dispersion
(Padoan \& Nordlund 1999; Mac Low et al.\ 1998; Stone, Gammie
\& Ostriker 1998)\footnote{Cho, Lazarian \& Vishniac (2002) (also
Lazarian, this conference)
have argued that ``imbalanced" turbulence, in which the turbulent
excitation is not statistically uniform in space, does not decay as
rapidly. This would correspond to the case where the turbulent driving
occurs at localized and discrete sites in space.
However, their numerical experiments were set up with rather
idealized initial conditions of an excess of traveling wave packets in 
one direction. The more realistic
simulations of Avila-Reese \& V\'azquez-Semadeni (2001), with
the initial turbulent driving applied to models of the ISM by discrete,
stellar-like sources, exhibit as rapid a decay as that of the
experiments mentioned earlier. Thus, so far most numerical 
evidence still points toward rapid decay of undriven turbulence.}. 

A possible driving mechanism for the turbulence in molecular 
clouds is dynamical (``bending mode") instabilities (Vishniac  
1994) in the compressed layers between converging flows, as in
the simulations of Hunter et al. (1986), Walder \& Folini (1998)
and Koyama \& Inutsuka (2002). It is important to note that
these instabilities appear to be enhanced by the presence of
cooling, so that the formation of turbulent molecular clouds
at temperatures of a few tens of Kelvins out of converging 
streams diffuse gas with $T>100$ K appears particularly likely.
Although this process has not been explicitely looked for in
multi-scale simulations of the ISM, the adaptive-grid, high 
resolution simulations presented by {\bf de Avillez} in
this session hold great promise for identifying it. In the meantime, his 
contribution discusses the relative values of the thermal and
magnetic pressures in the various temperature and density
regimes of the ISM. Other contributing driving mechanisms for the
turbulence may be the variation of the background UV radiation, which
induces mass flux among the warm and cool stable phases of the ISM
(Kritsuk \& Norman 2002) and the ``drag instability'' found and
discussed by {\bf Gu, Lin \& Vishniac}.

Within molecular clouds, the formation of their clumps and 
cores appears most likely a consequence of the supersonic turbulence
within the clouds, as it appears to have a dominant role in their
overall energy balance (corresponding to the standard notion that
molecular clouds are in general 
``supported" against their self-gravity by the turbulent 
kinetic energy). Moreover, since stars form in dense molecular cloud
cores, it is at this level of the fragmentation hierarchy that
we have reached the connection to star formation (SF). However,
if clumps and cores are density fluctuations produced by the
supersonic turbulence in molecular clouds, it is natural to ask
whether such a turbulent origin is 
compatible with the standard model of star formation (Mouschovias
1976; Shu 1977; Shu, Adams \&  Lizano 1987), in which clumps and cores
forming low mass stars are assumed to be 
in quasi-magnetostatic equilibrium, and evolving on long time scales,
given by the ambipolar diffusion characteristic time. This long
contraction time scale is at the basis of the low SF efficiency (SFE)
in the standard model. 

In recent years, a new model has been emerging,
which attempts to understand such issues as the properties of
molecular cloud cores, the efficiency of star formation, and even the
stellar initial mass function (IMF) as a consequence of the global
parameters of the turbulence in molecular clouds (Padoan 1995; Padoan
\& Nordlund 2002; \VS, \BP\ \& Klessen 2003; see also the
reviews by \VS\ et al. 2000, Mac Low \& Klessen 2003, and {\bf \BP}). In this
model, the low SFE is a consequence of {\it global} (large-scale)
turbulent support for the clouds, together with {\it local}
(small-scale) collapse induced by the compressible turbulence itself,
in a sort of turbulent ``colander''. Thus, only a small fraction of
the mass in a molecular cloud can proceed to gravitational collapse,
while the rest is probably dispersed in a few turbulent crossing
times, due to both stellar activity and the excess turbulent
energy. This latter process necessarily requires simulations of the
ISM at large, in which clouds can be adequately simulated as open
systems (see, e.g., the review by \VS\ 2002 and
references therein, and the contributions by {\bf de Avillez} and {\bf
Sarson et al.}). In particular, {\bf Sarson et al.} 
show promising results towards understanding the SF rate as a function 
of the mean gas density (the ``Schmidt law'') in the ISM directly from
numerical simulations. 

Thus, while in the standard model the 
low SFE is the result of a long evolution time scale for {\it each}
core, in the turbulent model it is a {\it statistical} issue, because
most of the mass in a molecular cloud is supported against collapse by
the turbulence, with most cores being transient, and only a small
fraction of the mass being in collapsing cores. It is a task of the
emerging theory to determine how this fraction is determined by the
global turbulent parameters. The contribution by {\bf
Ballesteros-Paredes} reviews the 
accomplishments of numerical simulations of MC turbulence in producing 
realistic cores, and the present state of the developing theory, as
well as some shortcomings. Further results of the turbulent model on
the resulting mass spectra of the collapsed object in turbulent clouds 
are given by {\bf Klessen}, on the basis of ``smoothed particle
hydrodynamics'' (SPH) numerical simulations. A promising numerical
code for extending SPH to the magnetic case is presented by {\bf Price 
\& Monaghan}.

Observationally, the contribution by {\bf Crutcher} makes an 
attempt at discerning between the two models,
concluding that {\it both} turbulence and magnetic fields are
important in MCs and SF, but finding that the presently availbale
evidence is still insufficient to discern between the competing models. The
contributions by {\bf Lai, Velusamy \& Langer} and by {\bf Wolf,
Launhardt \& Henning} present magnetic field determinations using
interesting novel techniques. As mentioned by {\bf Crutcher}, a large
database of this kind of measurements is needed in order to provide 
solid observational constraints as to which model better describes the 
star formation process as a part of the hierarchical fragmentation
sequence from the largest scales in the ISM.

All in all, the contents of this session reflect the exciting advances 
of recent years in both observations and theoretical understanding of
the hierarchical structure in the ISM and the SF process, and point
towards the necessary steps for the future.

\vskip -.3cm

\acknowledgements
Useful comments from J.\ Ballesteros-Paredes are gratefully
acknowledged. This work has received financial support from CONACYT
grant 36571-E.

\end{document}